# Simultaneous Multi-frequency Topological Edge Modes between One-dimensional Photonic Crystals

Ka Hei Choi,[1] C. W. Ling,[1] K. F. Lee,[1] Y. H. Tsang,[1] Kin Hung Fung[1],*

[1]*Department of Applied Physics, The Hong Kong Polytechnic University, Hong Kong, China*
*Corresponding author: khfung@polyu.edu.hk*

**We show theoretically that, in the limit of weak dispersion, one-dimensional (1D) binary centrosymmetric photonic crystals can support topological edge modes in all photonic band gaps. By analyzing their bulk band topology, these "harmonic" topological edge modes can be designed in a way that they exist at all photonic band gaps opened at the center of the Brillouin Zone, or at all gaps opened at the zone boundaries, or both. The results may suggest a new approach to achieve robust multi-frequency coupled modes for applications in nonlinear photonics, such as frequency up-conversion.**

*OCIS codes: (230.5298) Photonic Crystals; (350.1370) Berry's phase; (190.2620) Harmonic generation and mixing; (240.5590) Surface waves;*

The study of topological properties in condensed matter physics emerges from the discovery of quantum hall effect, where topological protected states exist at the edge of a crystal and are robust against small perturbations [1–6]. Recent studies have discovered that these topological properties exist also in photonic systems [7–13] and acoustic systems [14]. Rigorous mathematical treatments on making analogies between electronic and photonic systems have also been proposed by Lein and De Nittis [15–17].

As a geometric phase initially obtained from electron Bloch state, Zak phase is considered to be a $Z_2$ topological invariant ($\pi$ or 0) for classifying bands of one-dimensional electronic systems with inversion symmetry [18,19]. By making analogies, Zak phase can also be obtained from the photonic Bloch state of a centrosymmetric 1-D photonic crystal [20]. Zak phase describes the bulk band topology of the photonic band structure and determines the existence of edge modes. In photonic systems, it has been shown that Zak phase is also related to the reflection phase and surface impedance [20].

It is well-known that photonic crystals can be designed to confine light in small volumes for enhancement of light-matter interaction [21–25]. For example, by introducing a defect layer into a perfect 1-D photonic crystal, it can support a resonant state for confining light within the defect region [26–30]. This provides an opportunity to design devices for nonlinear photonic applications [31]. Instead of introducing a defect layer, recent discovery of topological properties in 1-D photonic crystal offers a new perspective to create resonant states (edge modes). Topological approach guarantees the creation of edge modes for a range of dielectric properties and structure of 1-D photonic crystals, which has the benefit of producing robust topological edge modes for enhancing light-matter interaction.

To enhance frequency conversion in nonlinear optics, it is important to enhance the light localization at both excitation and emission frequencies. Since topological edge modes can be supported in photonic crystals at more than one frequency, it could be useful for supporting enhancement at both excitation and emission frequencies. Based on the topological principles stated in Ref. [20], we further propose, in this Letter, that two combined 1-D photonic crystals can be specially designed to have bulk band topologies that guarantee the existence of "harmonic" topological edge modes under the limit of weak dispersion. We also show that it is able to choose whether these modes should exist at all band gaps opened at the center of the first Brillouin Zone (BZ), or at all band gaps opened at the zone boundaries, or both. These topologically protected edge modes at multiple frequencies could be robust against weak nonlinearity and localized at the same area, allowing additional enhancement of nonlinear effects such as frequency conversion.

We first consider a fundamental 1-D centrosymmetric photonic crystal (photonic crystal X in Fig. 1). It is created by stacking alternating layers of A and B with the normal of their planes parallel to the z-axis, which is also the direction of electromagnetic wave propagation. For a centrosymmetric unit cell, a single layer of A is sandwiched between half layer of B. The period ($\Lambda$) of an unit cell is given by $\Lambda = d_A + d_B$, where $d_A$ and $d_B$ are the thickness of a single dielectric layer of A and B. In this paper, the optical path length of each unit cell is specified by Eq. (1) for the purpose of creating a topological non-trivial 1-D photonic crystal [20],

$$\alpha \Lambda = n_A d_A + n_B d_B, \qquad (1)$$

where $n_A$ and $n_B$ are the refractive indexes of the non-dispersive dielectric layer A and B, $\alpha$ is the ratio of optical path length of a unit cell to its period. For the sake of simplicity, we first assume that $n_B = 1.5$ and every dielectric medium has negligible dispersion effect and magnetic response. Layer B is considered as the "reference" medium because its dielectric property will always remain unchanged in all of our simulations.

To understand how we produce topological edge modes periodically, we provide a brief review on the topological properties in a 1-D photonic crystal. For each photonic pass band, its topological phase is characterized by the quantized Zak phase of $\pi$ or 0, which is given by [20],

$$\theta_m^{\text{Zak}} = \int_{-\frac{\pi}{\Lambda}}^{\frac{\pi}{\Lambda}} \left[ i \int_{\text{unit cell}} \varepsilon(z) u_{m,K}^*(z) \partial_K u_{m,K}(z) dz \right] dK, \quad (2)$$

where $i \int_{\text{unit cell}} \varepsilon(z) u_{m,K}^*(z) \partial_K u_{m,K}(z) dz$ is the Berry connection, $\varepsilon(z)$ is the function of permittivity in space, and $u_{m,K}(z)$ represents the Bloch eigenfunction of electric field at the $m^{\text{th}}$ photonic pass band with a Bloch wavevector $K$.

Since the reflection phase of a 1-D photonic crystal is a physical observable for Zak phase, the existence of a topological edge mode can be predicted from finding the total sum of Zak phase below the $m^{\text{th}}$ band gap ($\sum_i^m \theta_i^{\text{Zak}}$) [20,32]. Suppose there are two 1-D photonic crystals, X and Y. When the sum $\sum_i^m \theta_i^{\text{Zak}}$ of X is different from that of Y by $(2l+1)\pi$, where $l$ is an integer, a boundary formed by combining X and Y will support topological edge mode(s) within this band gap [20]. The above X and Y can be found by changing the dielectric compositions of a photonic crystal continuously but keeping the optical path length of the unit cells unchanged [see Eq. (1)]. For example, by changing the refractive index $n_A$ of X continuously, some photonic band gaps of X will vary in size, while some band gaps may eventually close. A band crossing occurs when a band gap is closed and then re-opened under continuous change of refractive index. It is a topological phase transition of a photonic crystal since the Zak phases of the pass bands just below and above this gap changes by $\pi$ simultaneously. This important topological phenomenon will change $\sum_i^m \theta_i^{\text{Zak}}$ by $\pi$ [20]. This means that if the $n_A$ of X is tuned just before topological phase transition and the $n_C$ of Y is just after the transition, the combined structure (Fig. 1) will support a topological edge mode within this band crossing gap due to their $\pi$ difference in $\sum_i^m \theta_i^{\text{Zak}}$.

The principle above gives us a hint to create multiple numbers of topological edge modes simultaneously. It tells us that if there is a 1-D photonic crystal with multiple band gaps crossing simultaneously, we can design a photonic crystal interface which supports topological edge modes in all these crossing gaps. These photonic crystals with multiple band crossing can be found using the band crossing condition, $(n_A d_A)/(n_B d_B) = s_1/s_2$, where $s_1$ and $s_2$ are integers, and band crossing occurs at the $(s_1 + s_2)^{\text{th}}$ band gap [20].

The band crossing condition suggests that various combinations of photonic crystals of different parameters will have different types of "harmonic" topological edge modes. First, let us consider the case of $s_1 = s_2$ (i.e., $n_A d_A = n_B d_B$). As $s_1 = s_2$, $s_1 + s_2$ is always an even number, which suggests that when $n_A d_A = n_B d_B$, the number of band crossings is maximized and the crossings will all occur at every band gap that open at the center of BZ (central band gaps). To avoid confusion, this critical refractive index of $n_A$ at $s_1 = s_2$ is denoted as $n_\alpha$. From the condition $n_\alpha d_\alpha = n_B d_B$ and Eq. (1), it can be shown that $2/\alpha = (1/n_\alpha) + (1/n_B)$. As an example, we set $n_B = 1.5$ and $\alpha = 1.9$ for our 1-D photonic crystals, which gives $n_\alpha = 2.59$. Using the arguments above, topological edge modes can be generated at every central band gap by choosing some refractive indexes such that $n_A < n_\alpha$ and $n_C > n_\alpha$. In this case, the sum $\sum_i^m \theta_i^{\text{Zak}}$ for every central band gap of X is different from Y by $\pi$.

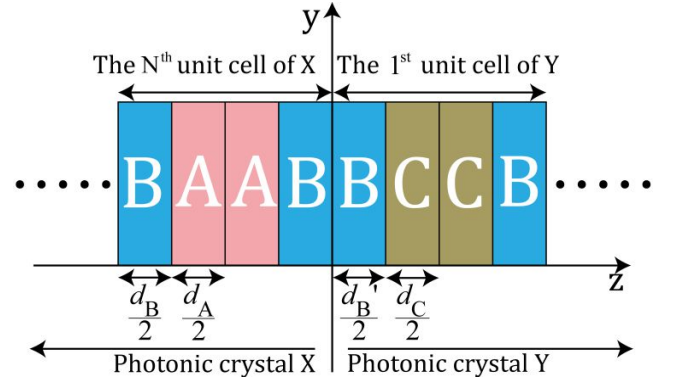

Fig. 1. Proposed 1-D photonic crystal model constructed by connecting two 1-D photonic crystals X and Y (X+Y), where X and Y contain $N$ unit cells. For both X and Y, their thickness is given by Eq. 1, and a single layer of A (or C) is sandwiched between half layer of B. Additionally, A and C has a small differences of refractive index.

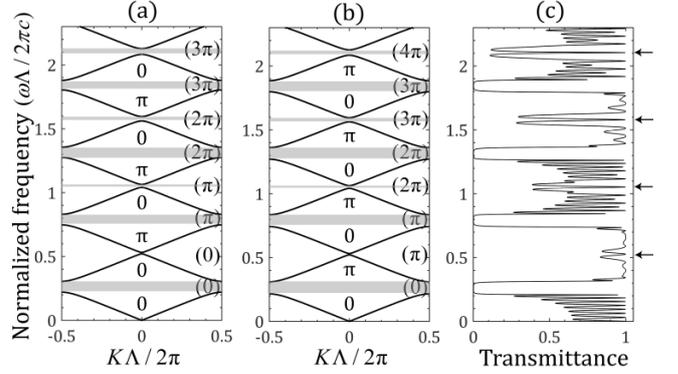

Fig. 2. (a) Photonic band structure of X, where $n_A = 2.51, n_B = 1.5$ and $\alpha = 1.9$. The Zak phase of each pass band is printed at the center of the pass band. Grey regions denote band gaps, with numbers in parentheses are the sum of the Zak phases below the band gap ($\sum_i^m \theta_i^{\text{Zak}}$). (b) Photonic band structure of Y, where $n_C = 2.67$, $n_B = 1.5$ and $\alpha = 1.9$. (c) Transmission spectrum of X+Y photonic crystals, where the number of units $N_X = 4$ and $N_Y = 4$. Black arrows indicate there are topological edge states.

Figs. 2(a) and 2(b) shows the photonic band structure of X and Y, which is plotted from the dispersion given by [33],

$$\cos(K\Lambda) = \cos(k_A d_A)\cos(k_B d_B) + \left(\frac{Z_A}{Z_B} + \frac{Z_B}{Z_A}\right)\sin(k_A d_A)\sin(k_B d_B), \quad (3)$$

where $K$ is the Bloch wavevector, $Z_i = (\mu_i/\varepsilon_i)^{\frac{1}{2}}$ is the impedance of the $i$-th medium. In the figures, we used relative permittivity instead of refractive index to describe the dielectric property of our 1-D photonic crystals. The critical refractive index $n_\alpha$ is represented in relative permittivity by $\varepsilon_\alpha = n_\alpha^2 = 6.71$ while the designed refractive indices are given by $\varepsilon_A = n_A^2 = 6.31$ and $\varepsilon_C = n_C^2 = 7.11$. Together with Eq. (3), the photonic band structures of X and Y are plotted in Figs. 2(a) and 2(b), respectively. Each photonic pass band is labelled by the Zak phase of its own, which is obtained from finding a special set of frequencies $\omega_Z$ by $\sin(\omega_Z n_B d_B/c) = 0$ [20], where $c$ is the speed of light. This special set of frequencies determines the Zak phases of the pass bands. When $\omega_Z$ is within a pass band, the Zak phase of this pass band is $\pi$, otherwise, it is 0 [20].

From the simulation results, the topological edge modes can be observed in the transmission spectrum and verified by the bulk band topology of X and Y. In Figs. 2(a) and 2(b), the Zak phase of each pass band is labelled at the center of their own band. The grey region represents photonic band gaps, while the numbers in parentheses on the right hand side are the $\sum_i^m \theta_i^{Zak}$ of the band gap. Due to the combination of these unique bulk band topologies (Figs. 2(a) and 2(b)), topological edge modes will all exist in zone-center band gaps and they are shown in the transmission spectrum of X+Y photonic crystals calculated using the Transfer Matrix Method [33] (See Fig. 2(c)). Theoretically, transmission should be low within the band gaps due to the absence of Bloch state. However, there are resonant transmission peaks (indicated by tiny black arrows) in some band gaps, which are due to the resonances associated with the edge modes. On the other hand, for photonic band gaps that open at the zone boundaries, $\sum_i^m \theta_i^{Zak}$ of these band gaps are all identical in Figs. 2(a) and 2(b), and thus there will be no resonance transmission peak exists within these band gaps, which is consistent with our previous discussions.

Other than adjusting dielectric properties of our 1-D photonic crystals, the unit cell structure also influences the corresponding Zak phase. We can manipulate the bulk band topology of photonic crystals by shifting the origins of their unit cells. A 1-D centrosymmetric photonic crystal with inversion symmetry always possesses of two inversion centers [18,20]. If the origin of a unit cell is shifted from one inversion center to the another one, the Zak phase of each photonic pass band will change from 0 to $\pi$ (or from $\pi$ to 0) [20]. Based on this approach, we can demonstrate that topological edge states can be created not only periodically at every zone-center band gap, but also periodically at every zone-boundary band gap.

Suppose there are two photonic crystals, X and X' [Fig. 3(b)], where both of them have the same dielectric compositions as X in Fig. 3(a) but different unit cell origins. Due to the shift of origin, every Zak phase in X will be different from X' by $\pi$. As a result, the bulk band topology of X' will be similar to Y [Fig. 2(b)], with an exception for the 0th pass band because the Zak phase of the 0th pass band is determined only by $\exp(i\theta_0^{Zak}) = \text{sgn}\left[1 - \frac{\varepsilon_A \mu_B}{\varepsilon_B \mu_A}\right]$ [20]. This additional Zak phase difference of $\pi$ between X and X' in the 0th pass band turns out to be significantly influential. We can conclude that when X and Y has a difference of $\pi$ in $\sum_i^m \theta_i^{Zak}$, X and X' will have a difference of $2\pi$ in $\sum_i^m \theta_i^{Zak}$; on the other hand, if X and Y have identical $\sum_i^m \theta_i^{Zak}$, X and X' will have a difference of $\pi$ in $\sum_i^m \theta_i^{Zak}$. This implies that shifting the origin of the unit cell without changing its dielectric properties can produce periodical topological edge modes at every zone-boundary band gap. Whenever there is a topological edge mode in the zone-center band gap of X+Y, the mode will be absent from the zone-center band gap of X+X'. On the opposite side, as topological edge modes are absence from the zone-boundary band gaps of X+Y, these modes will appear in the zone-boundary gap of X+X'. This phenomenon is shown in the transmission spectrum of X+X' in Fig. 3(b). The resonant transmission peaks resulted from topological edge modes are found to exist periodically at each edge band gap instead of the central band gap, which are indicated by black arrows.

Another significant consequence due to the Zak phase change of $\pi$ at the 0th pass band is the creation of periodical topological edge mode at both zone center gaps and zone boundary gaps. This can be done by considering another set of 1-D photonic crystals, X and Y' [Fig. 3(c)]. Our 1-D photonic crystal Y' has combined both modifications we discussed previously, where Y' have the same dielectric property of Y, and Y' shifts its origin from one inversion to another as if X'. Now, consider the Zak phase of Y' after both modifications: With the dielectric property of Y, the Zak phase of each pass band will be different from X by $\pi$ except the 0th pass band. Then, shifting the origin of Y from one inversion center to another will change the Zak phase of each pass band by $\pi$ again, but this time including the 0th pass band. As a result, except the 0th pass band, the Zak phase of each pass band in Y' has returned to its initial phase because they change by $\pi$ twice, i.e. $2\pi$. However, at the 0th pass band, the Zak phase only changes by $\pi$ once because the modification of dielectric properties does not influence the Zak phase at the 0th pass band. Accordingly, for X and Y', the differences of $\sum_i^m \theta_i^{Zak}$ of every band gap are $3\pi$, excepted that the first band gap has the differences of $\pi$ only. As we have described before, this implied that topological edge modes will exist at every zone-center and zone-boundary band gaps. We can find these topological edge modes in the transmission spectrum of X+Y' in Fig. 3(f), where resonant transmission peaks are observed in both zone-center and zone-boundary band gaps due to the combination of their unique bulk band topologies.

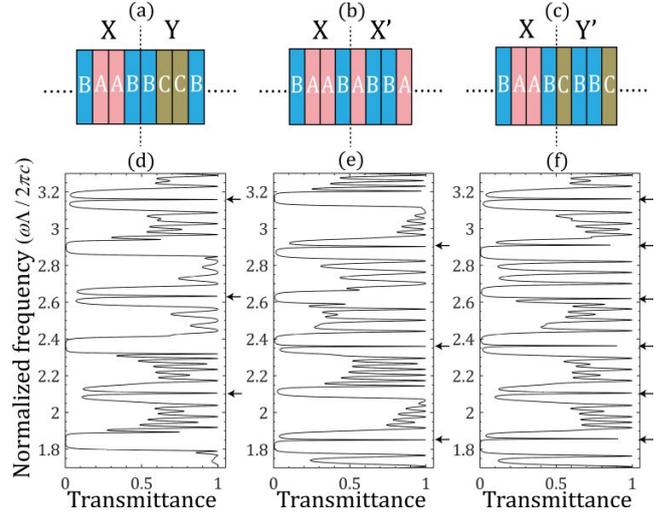

Fig. 3. (a) The photonic crystal structure of X + Y, (b) X + X', and (c) X + Y'. (d) The transmission spectrum of X + Y, (e) X + X', and (f) X + Y'. Each photonic crystals above are composed by stacking 4 unit cells. From our basic assumption, the refractive index of layer B is given by $n_B = 1.5$ and $\alpha = 1.9$. The refractive index of layer A of photonic crystal X is given by $n_A = 2.51$, and the refractive index of the layer C for photonic crystal Y is $n_C = 2.67$. The topological edge modes of each band gap are also indicated by black arrows.

From our previous discussions, it seems that creating periodical topological edge modes in X+Y requires the freedom of dielectric properties. However, what it requires are only two materials with close refractive indexes ($n_A - n_C$ is small when compared to $n_\alpha$ itself) and the design of thicknesses. Suppose that our 1-D photonic crystals in Fig. 1 has no flexibility of choosing their dielectric properties, such that $n_B, n_A$, and $n_C$ are fixed. To design X+Y with periodical topological edge modes, we know that the thickness of our photonic crystals must be given by Eq. (1). As $n_A - n_C$ is small, we can assume that the critical topological phase transition point $n_\alpha$ is the average of $n_A$ and $n_C$. With $n_\alpha$ and $n_B$, we can obtain $\alpha$ by the relation $2/\alpha = (1/n_\alpha) + (1/n_B)$, and hence we know the thickness $d_A, d_C, d_B,$ and $d'_B$ for creating periodical topological edge modes at central band gaps. Therefore, as long as $n_A - n_C$ is small, periodical topological edge modes at central band gap can always be produced by thickness adjustment.

Nevertheless, it should also be noted that the value of $n_A - n_C$ determines an upper frequency limits of periodical topological edge modes. These modes are not guaranteed at high frequencies because small change of dielectric property can induce rapid change of the bulk band topology at high frequencies. Still, the upper frequency limit is pretty high for application purposes, which is at $\omega\Lambda/2\pi c \sim 5$ for our case and can be extended by reducing the value of $n_A - n_C$.

The simultaneous topological edge modes proposed in this Letter may be used to enhance nonlinear light–matter interactions such as frequency up-conversion. Some nonlinear optical applications utilizing localized defect modes in 1D photonic crystal have been proposed [21–25]. While previous works usually consider enhancement at single-frequency such as excitation frequency, it should be noted that enhancement at emission frequency also plays an important role. To demonstrate how the multi-frequency topological edge modes can be used to enhance nonlinear optical processes, we show the normalized intensity pattern inside the photonic crystal of X+Y' in Fig. 4. First, we note that, for every edge mode, there are strongly localized fields near the interface between X and Y'. The enhancement at the edge mode frequencies in Fig. 4 are about 20 to 30 while the enhancement region has a two-dimensional (2D) surface area of beam size of the incident light in plane wave approximation. The enhancement can increase exponentially with increasing number of units. This result suggests that the interface region is suitable for inserting single or few layers of thin or 2D materials for enhancing frequency up-conversion optical processes.

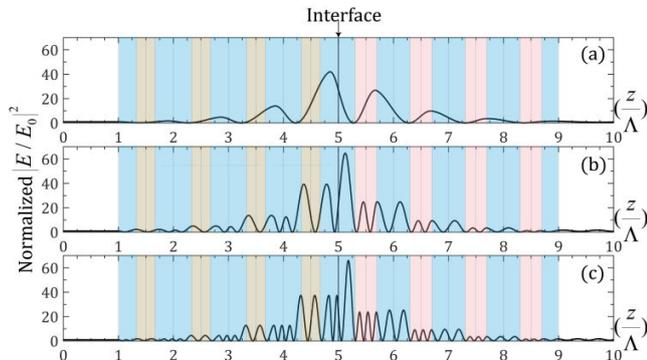

Fig. 4. Intensity profile for plane-wave excited topological edge mode in X + Y' photonic crystals, where the $|E|$ is normalized by the amplitude of the incident wave, $|E_0|$. A plane wave enters from the right hand side of the photonic crystal. The colored regions represent the dielectric composition of X and Y', while the uncolored regions are an air medium. It shows the topological edge modes at (a) the 1st, (b) the 2nd, and (c) the 3rd zone-boundary band gap.

In conclusion, we showed how 1-D binary photonic crystals can be designed to create topological edge modes in any photonic band gap by manipulating their bulk band topology. We have designed a photonic crystal interface supporting periodical topological edge modes that exist at all zone-center gaps, or at all zone-boundary gaps, or both. The edge modes are topologically robust and could be maintained in weak perturbations such as weak nonlinearities. The result may suggest a new approach to achieve robust multi-frequency coupled modes for applications in nonlinear photonics, such as frequency up-conversion, through simultaneous enhancement at both excitation and emission frequencies.

**Funding.** Hong Kong Research Grant Council (GRF project no. 15300315); The Hong Kong Polytechnic University (G-UA95).

**Acknowledgment.** The authors thank Raymond Wu, Tsz Chung Mok, Xianglong Miao, and Kam Tuen Law for fruitful discussions.